\PassOptionsToPackage{numbers, compress}{natbib}
\documentclass{article}
\usepackage[final]{neurips_2025}
\usepackage{graphicx}
\usepackage{dcolumn}
\usepackage{bm}

\usepackage{multirow}%
\usepackage{amsmath,amssymb,amsfonts}%
\usepackage{amsthm}%
\usepackage{mathrsfs}%
\usepackage[title]{appendix}%
\usepackage{xcolor}%
\usepackage{textcomp}%

\usepackage{booktabs}%
\usepackage{algorithm}%
\usepackage{algorithmicx}%
\usepackage{algpseudocode}%
\usepackage{listings}%

\usepackage[utf8]{inputenc} 
\usepackage{relsize}
\usepackage{textgreek}
\usepackage{textcomp}
\usepackage{siunitx}

\PassOptionsToPackage{hyphens}{url}\usepackage{hyperref}

\newcommand{\EFB}[1]{{$\text{EF}^B_{\text{#1}}$}}

\DeclareUnicodeCharacter{03BC}{\textmu}

\newif\iflong
\longtrue

\begin{document}


\title{Look mom, no experimental data! Learning to score protein-ligand interactions from simulations}





\author{%
Michael Brocidiacono$^1$ \quad James Wellnitz \quad Konstantin I. Popov \quad Alexander Tropsha \\
Eshelman School of Pharmacy \\
University of North Carolina at Chapel Hill \\
$^1$\texttt{mixarcid@unc.edu}
}


\maketitle

\begin{abstract}
Despite recent advances in protein-ligand structure prediction, deep learning methods remain limited in their ability to accurately predict binding affinities, particularly for novel protein targets dissimilar from the training set. In contrast, physics-based binding free energy calculations offer high accuracy across chemical space but are computationally prohibitive for large-scale screening. We propose a hybrid approach that approximates the accuracy of physics-based methods by training target-specific neural networks on molecular dynamics simulations of the protein in complex with random small molecules. Our method uses force matching to learn an implicit free energy landscape of ligand binding for each target. Evaluated on six proteins, our approach achieves competitive virtual screening performance using 100-500 μs of MD simulations per target. Notably, this approach achieves state-of-the-art early enrichment when using the true pose for active compounds. These results highlight the potential of physics-informed learning for virtual screening on novel targets. We publicly release the code for this paper at \url{https://github.com/molecularmodelinglab/lfm} under the MIT license.
\end{abstract}

\section{Introduction}

Finding small molecule ligands that bind to a protein target is a critical step in early-stage drug discovery. However, this objective can be extremely challenging and expensive -- the typical approach is to run a high-throughput screen (HTS), which can take months to set up and cost millions of dollars \cite{davies_streamlining_2006}. Identifying true binders faster and more cheaply is a major goal of computational drug discovery. 

Traditionally, the challenge of predicting protein-ligand binding affinities is separated into two tasks: docking and scoring. Docking algorithms propose 3D binding poses of a ligand within a protein’s binding site, while scoring functions estimate the binding affinity of a given pose. While much recent progress has been made in the area of docking \cite{corso_diffdock_2022, abramson_accurate_2024, wohlwend_boltz-1_2024}, scoring has lagged behind \cite{yang_predicting_2020, volkov_frustration_2022, brocidiacono_improved_2024}. This is the primary reason for the unreliability of structure-based virtual screening (SBVS), wherein ligands from a large library are ranked according to their predicted binding affinity, and the top-scoring hits are chosen for experimental characterization. More recently, generative modeling has emerged as a potential alternative to SBVS; however, generative models are still typically evaluated and/or fine-tuned according to inaccurate scoring functions such as AutoDock Vina \cite{trott_autodock_2010, peng_pocket2mol_2022, newman_structure_2021}.

The difficulty of scoring arises from both data limitations and fundamental trade-offs in existing approaches. ML-based scoring functions often suffer from overfitting due to the scarcity of high-quality binding affinity data \cite{yang_predicting_2020, volkov_frustration_2022}. On the other hand, physics-based methods exhibit a speed-accuracy trade-off: rigorous absolute binding free energy (ABFE) simulations provide accurate affinity estimates but require extensive computational resources, often taking many hours per molecule even on GPUs \cite{alibay_evaluating_2022, fu_meta-analysis_2022, aldeghi_accurate_2015}. In contrast, traditional fast scoring functions, such as those used by Vina and Glide \cite{friesner_glide_2004}, offer rapid predictions but are often unreliable, failing to achieve robust enrichment in virtual screening tasks \cite{brocidiacono_improved_2024, kolb_docking_2009, scior_recognizing_2012}.

Developing scoring methods that generalize beyond the available training data is especially important for the next generation of drugs. Recent advances in drug discovery have expanded interest beyond the traditional druggable proteome, targeting intrinsically disordered proteins \cite{saurabh_fuzzy_2023}, protein-protein interactions \cite{nada_new_2024}, and RNA tertiary structures \cite{childs-disney_targeting_2022} as promising avenues for next-generation therapeutics. The ability to efficiently identify small-molecule binders for these novel targets could greatly accelerate the development of new treatments. These novel target classes represent a formidable yet rewarding challenge for screening methods, as no current methodology can reliably find binders for such targets.

In this work, we propose combining the advantages of physics-based and machine learning models for protein-ligand scoring. Instead of trying to train a general binding affinity predictor to work on all protein targets, we seek to train a new model for each protein target of interest. We create a custom training set for each protein by running molecular dynamics (MD) simulations of random small molecules inside the binding site. We then use force matching to train a neural network on the free energy landscape of ligand binding. We emphasize that this approach requires no experimental data for the target in question except for its 3D structure (which can be estimated using structure-prediction methods such as AlphaFold \cite{jumper_highly_2021}). We name this workflow ligand force matching (LFM).

\begin{figure}
    \centering
    \includegraphics[width=0.80\textwidth]{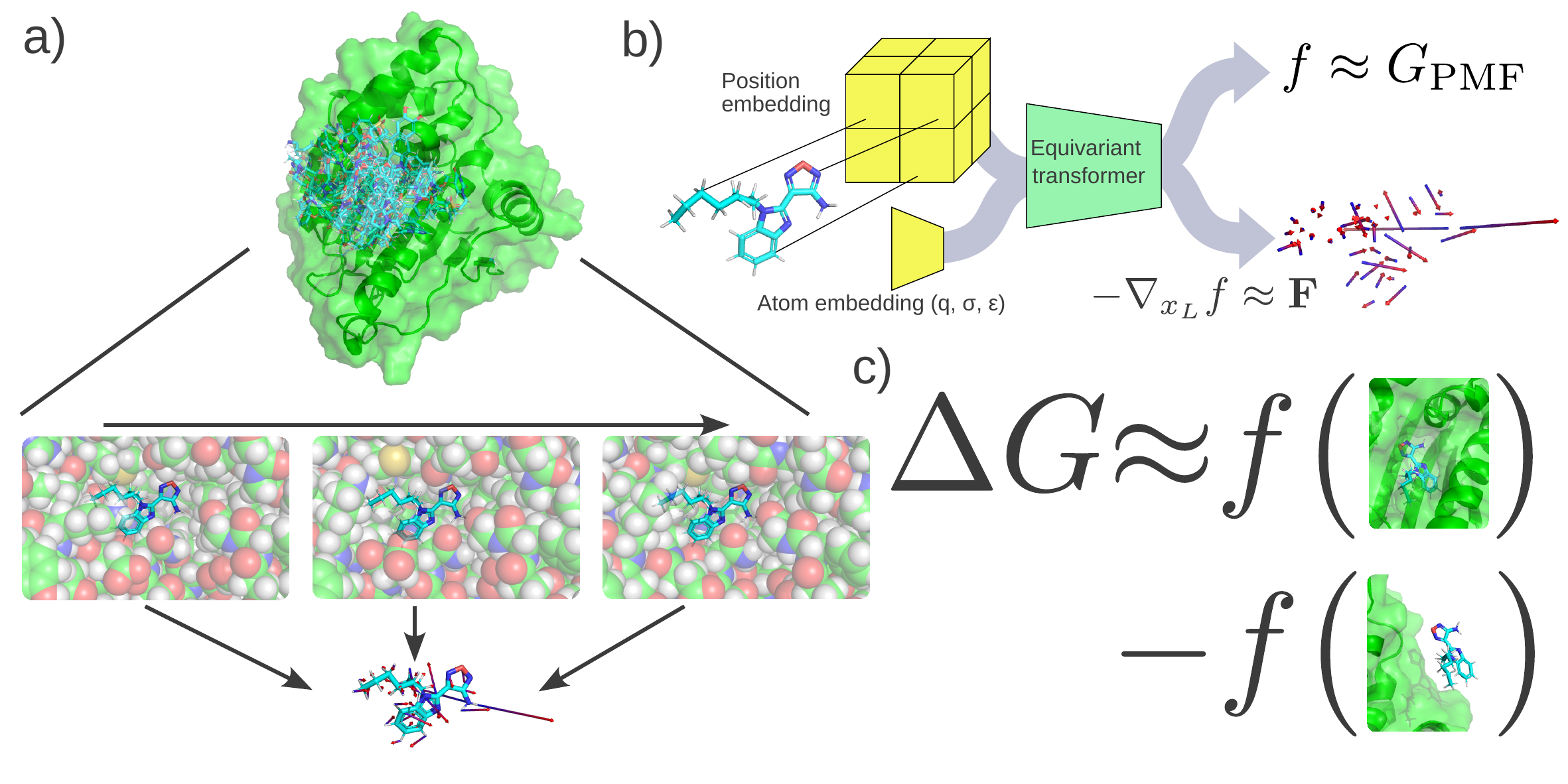}
    \caption{Method overview. a) For a particular protein target, we start by placing random small molecules in the binding site. For each ligand, we run an MD simulation of the protein-ligand complex, keeping the ligand atoms frozen relative to the binding site. We record the positions and intermolecular forces of the ligand atoms. b) Using this dataset, we train a neural network to approximate this binding PMF of the protein-ligand complex using force matching. c) In a virtual screen, we use this neural network to approximate the free energy of protein-ligand binding by subtracting the PMF of the ligand outside the binding site from the PMF of the docked ligand.}
    \label{fig:pmf_overview}
\end{figure}

\section{Method}

\subsection{Overview}

Invoking the Born-Oppenheimer approximation \cite{born_zur_1927}, a receptor-ligand system can be specified by the positions of the receptor atoms: $x_R$, the ligand atoms $x_L$, and the solvent atoms $x_S$. Throughout this paper, we use a coordinate system defined by the rigid-body alignment of the binding site $\alpha$ carbons to a reference structure. At equilibrium, these positions follow the Boltzmann distribution:

\begin{equation}
    p(x_R, x_L, x_S) = \frac{1}{Z} e^{-\beta U(x_R, x_L, x_S)}
\end{equation}

Here $U$ is the potential energy of the system and $\beta = \frac{1}{k_B T}$ is the Boltzmann factor. In practice, we can approximate $U$ with a fast and easily-evaluated sum of nonbonded and bonded energy terms.

When scoring protein-ligand complexes, we desire to predict the \textit{binding free energy} $\Delta G_\text{binding} = G_\text{bound} - G_\text{unbound}$. The free energy $G$ for a particular state is defined:

\begin{equation}
    G_\Omega = -\frac{1}{\beta} \log \int_\Omega e^{-\beta U(x_L, x_R, x_S)} dx_L dx_R dx_S
\end{equation}

Here $\Omega$ is the set of either all bound or all unbound poses, depending on the free energy in question.


The integral in this equation is computationally intractable, as protein-ligand systems can easily exceed 10K atoms. In practice, we can approximate $\Delta G_\text{binding}$ by sampling from many intermediate systems wherein the energy terms corresponding to the protein-ligand interactions are slowly attenuated; the final free energy is the work done during this transformation. This is the approach taken by ABFE protocols such as double decoupling \cite{gilson_statistical-thermodynamic_1997} and the alchemical transfer method \cite{wu_alchemical_2021}. As mentioned above, such algorithms are accurate but slow.

As an alternative approach, we can condition $G$ on the ligand coordinates $x_L$. This results in the \textit{binding potential of mean force (PMF).}

\begin{equation}
    G_\text{PMF}(x_L) = -\frac{1}{\beta} \log \int_{-\infty}^\infty e^{-\beta U(x_L, x_R, x_S)} dx_R dx_S
\end{equation}

As its name implies, the force this potential generates on the ligand atoms is equal to the ensemble average of the ligand forces from the full potential $U$.

\begin{equation}
    -\nabla_{x_L}G_\text{PMF}(x_L) = -\Bigl\langle\nabla_{x_L}U(x_L, x_R, x_S)\Bigr\rangle_{x_R, x_S}
\end{equation}

Estimating the full $\Delta G_\text{binding}$ is much easier if we have access to $G_\text{PMF}$, as it already integrates over the majority of the degrees of freedom. We can additionally quickly approximate the full free energy difference as the difference between $G_\text{PMF}$ of a single bound and unbound ligand pose.

While $G_\text{PMF}$ is still intractable to compute directly, we can easily estimate the forces $\mathbf{F} = -\nabla_{x_L}G_\text{PMF}$ via a single MD simulation.

In standard molecular mechanics force fields, $U$ can be split into the ligand-ligand and ligand-environment components $U(x_L, x_R, x_S) = U_{LL}(x_L) + U_{LE}(x_L, x_R, x_S)$. $U_{LL}(x_L)$ is easy to calculate, so, in practice, we want to estimate the intermolecular PMF $\hat{G}_\text{PMF} = G_\text{PMF} - U_{LL}$, utilizing the intermolecular forces $\mathbf{\hat{F}}$.

We desire a model that can quickly approximate $\hat{G}_\text{PMF}$; this should be able to generalize to new ligands but is specific to the current target in question. We do this by generating a dataset $\mathcal{D} = \left( x_L, a_L, \mathbf{F}\right)$ of random small molecules in random positions across the binding site. Here, $a_L$ are nonbonded parameters $[\sigma, \epsilon, q]$ needed to specify the Lennard-Jones and Coulombic energies for the force field.

Finally, we train a neural network $f$ to approximate $\hat{G}_\text{PMF}$ (up to a constant) by minimizing the force-matching loss \cite{wang_machine_2019}:

\begin{equation}
    \mathcal{L}_{F} = \bigl(-\nabla_{x_L} f(x_L, a_L) - \mathbf{\hat{F}} \bigr)^2
\end{equation}

This loss function has previously been used to train coarse-grained and implicit solvent force fields \cite{coste_developing_2024, katzberger_general_2024}.

When using the trained model in practice, we only want to evaluate the model on ``reasonable'' poses given to it by the docking program (that are presumably relatively low-energy), as well as poses outside the binding site (to subtract off the PMF in solvent). Therefore, when generating data, we only need to sample such reasonable poses, while ensuring that we sample along a path between the center of the binding site and somewhere outside. We describe one such method below.


\subsection{Data generation}

We begin by running a 100 ns MD simulation of the target receptor to produce an ensemble of conformations. We then compute the ``exit point'' $x_E$, defined as a point close to the pocket center that is at least $c$ (here, 9\AA) away from any receptor atom. We assume that a molecule centered at this exit point is far enough away from the receptor that its interactions are negligible. We then generate each datapoint as follows:

We sample a random ligand from the ZINC database \cite{sterling_zinc_2015}. We select a pose for this ligand by first generating a conformer with RDKit \cite{noauthor_rdkit_nodate}. We then uniformly sample a point $x$ between the pocket center and the exit point. We then place the center of mass of the ligand according to a normal distribution centered at $x$ and with standard deviation $\sigma_P$ (here 6 \AA). The ligand is rotated randomly around this position.

We then select a random apo protein conformation (with solvent) from its MD trajectory. The ligand is placed in the pocket according to the position specified above. The ligand now likely has steric clashes with the protein and solvent. Additionally, the ligand is likely in an unreasonable pose that will not help training. To rectify these two problems, we run a short (100 ps) MD simulation of the complex, slowly adding the interaction terms between the ligand and its environment according to a soft-core alchemical potential \cite{beutler_avoiding_1994} defined by the sterics interaction term $\lambda_s$ and the electrostatics term $\lambda_c$. We first drive $\lambda_s$ from 0 to 1 over 90 ps and then drive $\lambda_c$ from 0 to 1 over the remaining 10 ps. Additionally, we ensure that the center of mass of the ligand remains close to its starting position via a harmonic restraint (k= \SI{10}{\kilo\joule\per\mol\per\nano\metre\squared}). Thus, by the end of this alchemical simulation, all steric clashes should be removed. Additionally, the ligand is now in a more energetically favorable conformation while its center of mass is still in the relatively same position.

Now that we have a ligand position that meets our criteria, we run the final simulation for 4 ns to determine the mean force on the ligand atoms. During this simulation, we freeze the ligand atoms. Additionally, to ensure that the ligand remains in the same place relative to the protein binding site, we restrain the protein atoms using a harmonic force between the pocket atoms and the reference pocket atoms, projected so that it only affects the target's rigid-body degrees of freedom (see Appendix \ref{app:rigid}).

During the simulation, we log the intramolecular forces acting on the ligand atoms every 10 ps. We compute the final mean force after the equilibration time (here, 2.5 ns).

All data in this paper was generated via OpenMM \cite{eastman_openmm_2024} with the Amber ff14SB \cite{maier_ff14sb_2015} and Espaloma \cite{wang_end--end_2022} force fields.

\subsection{Model Architecture and Training}

To predict the PMF, our model needs to have embeddings for both the ligand atom features and the position of the atoms within the binding site. Note that this task is not invariant with respect to the ligand coordinates, so a purely equivariant architecture will not work.

To create the ligand atom embedding $h_L$, we utilize an MLP over the atom parameters $a$. The ligand coordinate embedding is done via cubic spline interpolation over uniformly spaced 3D grids. We define several such grids with various spacings and hidden dimensions. We use \texttt{torch-cubic-spline-grids} \cite{noauthor_teamtomotorch-cubic-spline-grids_2025} to define an embedding for each atom over each grid. The final ligand coordinate embedding is an MLP over the concatenated grid embeddings for each atom.

The final combined embedding $h_c$ is an MLP over the concatenated ligand embeddings, coordinate embeddings, and the multiplication of the two embeddings.

These embeddings are then fed into an equivariant transformer from TorchMDNet \cite{tholke_torchmd-net_2022}. The final energy is the sum of individual atom energies.

We train the model with force matching. In addition to just the force-matching loss, we also include the mean-squared-error loss for the force on the ligand center of mass $\mathcal{L}_{\bar{F}}$ and torque of the ligand center of mass $\mathcal{L}_{\bar{T}} $. This is done because we care more about the relative PMFs of the ligand in different positions. The final loss is thus $\mathcal{L} = \lambda_F\mathcal{L}_{F} + \lambda_{\bar{F}}\mathcal{L}_{\bar{F}} + \lambda_{\bar{T}}\mathcal{L}_{\bar{T}}$. In this paper, we use $\lambda_F = 1.0$, $\lambda_{\bar{F}} = 50.0$, and $\lambda_{\bar{T}} = 5.0$.

\subsection{Model use}


To score a docked ligand, we first compute the total energy (intramolecular energy + model energy $f(x)$) for each docked pose. We select the minimum energy pose and compute the final score as $\Delta G_\text{binding} \approx f(x) - f(x_\text{solv})$. Here, $x_\text{solv}$ is the ligand outside the binding site. (To compute this, we re-center the ligand at the exit point defined during data generation.) 

This approximation of binding affinity assumes a rigid ligand and therefore neglects the ligand reorganization energy of binding. A more accurate (albeit slower) approximation would be to explicitly compute the free energy between the solvated and bound structures, e.g., by using the alchemical transfer method \cite{wu_alchemical_2021} over the PMF. We leave this to future work.

\section{Results}

\subsection{BRD4 virtual screening}

We first tried this method on the well-studied cancer target BRD4, utilizing an older variant of the dataset generation (outlined in Appendix~\ref{app:vs}) to generate ~50K datapoints. After achieving promising results on retrospective data (shown below), we used our model to screen a 170K in-house library and selected the top-predicted 
binders for experimental validation.

\begin{figure}
    \centering
    \includegraphics[width=0.9\columnwidth]{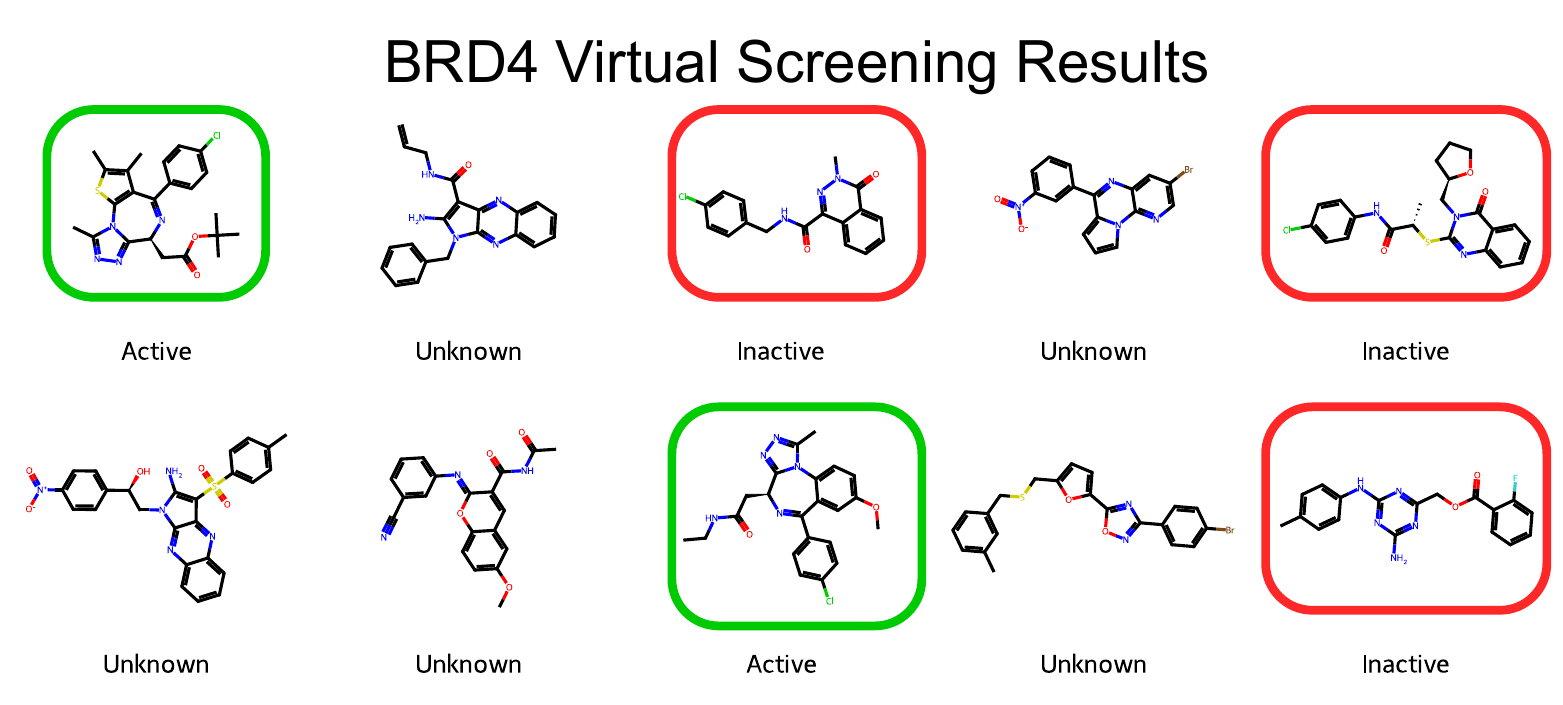}
    \caption{Results from the in-house virtual screen with BRD4. We could not find 5 of the compounds in the collection, so they were not tested; of the three that were tested, none were found to bind.}
    \label{fig:brd4_hits}
\end{figure}

As seen in Figure~\ref{fig:brd4_hits}, two of the top ten compounds selected are known BRD4 binders. Encouraged, we elected to test the remaining 8 for BRD4 binding via Isothermal Calorimetry (ITC). Unfortunately, we were only able to acquire 3 of the compounds from the library; of these, none demonstrated measurable binding affinity. Thus, though we failed to discover novel binders, we did rediscover two known BRD4 binders.

\subsection{Benchmarking on all targets}

We additionally tried this method on 5 other targets (MCL1, ESR1, MDM2, CDK2, and HSP90). We chose these targets for a combination of factors, namely their small size (enabling faster and cheaper MD simulations) and their large and diverse number of known binders, both with and without co-crystal structures (enabling us to robustly test virtual screening performance).

\begin{figure}
    \centering
    \includegraphics[width=1.0\textwidth]{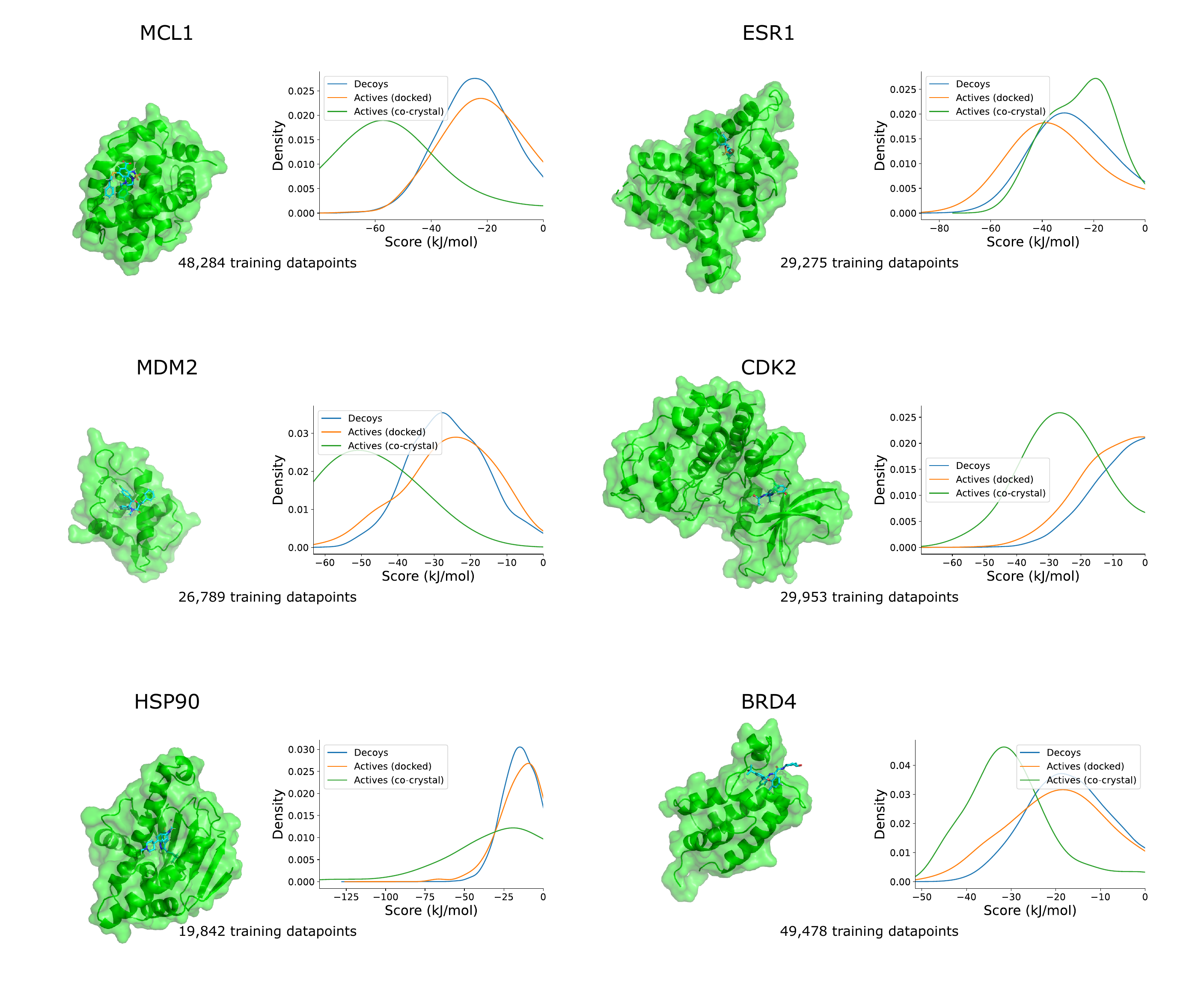}
    \caption{Each benchmarking target used in this study, co-crystallized with a representative ligand. For each target, we show the distribution of LFM scores for actives and decoys docked with \textsc{DiffDock}, as well as the co-crystal structures of the actives with known structure. In general, the scores of the docked actives are lower than the docked decoys, showcasing the value of the LFM models for virtual screening. We see a more dramatic score shift for the co-crystallized active molecules, showcasing the pose-dependence of the LFM models. }
    \label{fig:score_dist}
\end{figure}

For each protein, we compiled a set of protein-ligand cocrystal structures from the PLINDER \cite{durairaj_plinder_2024} dataset. We chose an arbitrary structure to initialize our method from this set, along with a set of aligned ligands with 3D poses. We also searched ChEMBL \cite{gaulton_chembl_2012} for known binders and selected a diverse subset of small molecules with known activity but no known binding pose. Finally, we generated up to 50 property-matched decoys for each active molecule, following the example of DUD-E \cite{mysinger_directory_2012}. Our primary evaluation metric is the ability of LFM to distinguish known binders from these decoys, ideally with high early enrichment.

We ran the LFM workflow for each protein, ensuring that no molecule used for training the model was within 0.3 Tanimoto similarity coefficient to any known binder from ChEMBL (2048-bit, radius 4 Morgan fingerprint \cite{morgan_generation_1965}). This enables us to robustly test the ability of our method to generalize across chemical space. We generated 25K-60K datapoints per target, utilizing the cloud resources of VastAI \cite{noauthor_rent_nodate}. We split the dataset into 80:10:10 train:validation:test sets. This amounts to 100-500 μs of training MD simulation time per protein, for a cost of around \$1K each. We trained each model for up to 300 epochs on an NVIDIA L40 GPU, saving the model that achieved the best validation mean force loss. We used the AdamW \cite{loshchilov_decoupled_2019} optimizer with a learning rate of $10^{-4}$.

None of these targets is especially novel, meaning that traditional screening methods are likely to work adequately well on them already. We would ideally like to test our method on targets that are out-of-domain for current methods. However, by their nature, novel targets lack large numbers of diverse binders that can robustly allow us to test the performance of this method retrospectively. Thus, we leave the evaluation of LFM on novel target classes to future prospective studies.

We used both Uni-Dock \cite{yu_uni-dock_2023} and \textsc{DiffDock-L} \cite{corso_deep_2024, corso_diffdock_2022} to dock the binders and decoys for each target. Uni-Dock implements a faster GPU-enabled version of the algorithm used by Vina. \textsc{DiffDock} has demonstrated improved docking accuracy relative to traditional methods such as Vina. It should perform especially well on these well-studied targets since it was trained on all of them. The use of these two docking methods will demonstrate how improved docking performance impacts the performance of scoring models.

We then used the LFM models, along with the baseline scoring models Vina \cite{trott_autodock_2010} and \textsc{Gnina} \cite{mcnutt_gnina_2021} to score and rerank the resulting poses. \textsc{Gnina} is a 3D CNN-based model that has previously demonstrated state-of-the-art performance in virtual screening \cite{sunseri_virtual_2021}. Since its neural networks were trained using 3-fold cross-validation with clustered splits \cite{francoeur_three-dimensional_2020}, we used an ensemble of neural networks that were not trained on the cluster containing the target in question.


The primary metric for evaluating the scoring models is the Bayes enrichment factor \EFB{} \cite{brocidiacono_improved_2024}. Given a cutoff fraction $\chi$ and a set of model scores $S$ on both active and decoy compounds, this metric is an estimate of the probability ratio $\frac{P(\text{Active}|S>S_{\chi})}{P(\text{Active})}$. This is the probability that the top-scoring hits from a virtual screen will bind, relative to selecting compounds at random. We are primarily interested in the maximum achievable enrichment \EFB{max}, though we also report the enrichment at $\chi = 1\%$. It is worth noting that, since these metrics depend so heavily on the tail of the score distribution, they are highly sensitive to noise in the data. Thus, we should pay close attention to the (asymmetric) error bars (computed via bootstrapping).

\subsection{Benchmarking results}

As shown in Table \ref{tab:med_crystal_metrics}, the LFM models perform better along all metrics when utilizing true crystal poses for the actives molecules and either Uni-Dock or \textsc{DiffDock} for the decoy molecules. Additionally, the LFM models can more accurately pick out ``correct'' docked structures (within 2 \AA RMSD) from a set of proposed structures.

The story becomes more complicated when we use docked poses for both the actives and decoys. As we see in Table~\ref{tab:med_metrics}, the LFM models perform about as well in terms of enrichment as the other models when using both Uni-Dock or \textsc{DiffDock} for both the actives and decoys. Additionally, \textsc{Gnina} is the best at classifying actives from inactives in terms of AUC.

We suggest that these results can be explained by the increased pose-sensitivity of the LFM models. The score given by \textsc{Gnina} (here, the \texttt{CNN\_VS} output) is the product of an affinity score and a pose score; thus, it is more likely to rank an active above a decoy even if both their poses are deemed incorrect. This especially explains the higher AUCs achieved by \textsc{Gnina} relative to its enrichment factors, as these incorrectly docked actives will not be in the top selected hits despite contributing to AUC.  

\begin{table}
\centering
\caption{Median metrics for each model when using known co-crystal poses for the active compounds and using Uni-Dock or \textsc{DiffDock} to dock the decoy compounds. We additionally report docking accuracy at a 2 \AA\ cutoff when using the model to rerank the poses proposed by Uni-Dock or \textsc{DiffDock}.}
\label{tab:med_crystal_metrics}
\begin{tabular}{lllllll}
\toprule
Docking method & Model & \EFB{max} & \EFB{1\%} & AUC & $\% < 2$ \AA \\
\midrule

\multirow{3}{*}{Uni-Dock} &Vina & 3.7 [1.0, 13] & 0.0 [0.0, 0.0] & 0.36 \textpm \ 0.09 & 13 \textpm \ 5.0 \% \\
&\textsc{Gnina} & 120 [16, 230] & 12 [5.9, 23] & 0.79 \textpm \ 0.05 & 13 \textpm \ 5.4 \% \\
&LFM & \textbf{290 [120, 620]} & \textbf{28 [18, 44]} & \textbf{0.88 \textpm \ 0.07} & \textbf{19 \textpm \ 5.2 \%} \\
\midrule

\multirow{3}{*}{\textsc{DiffDock}} &Vina & 25 [12, 72] & 8.5 [4.3, 22] & 0.69 \textpm \ 0.06 & 63 \textpm \ 6.8 \% \\
&\textsc{Gnina} & 120 [28, 230] & 20 [11, 28] & 0.83 \textpm \ 0.06 & 56 \textpm \ 6.9 \% \\
&LFM & \textbf{640 [68, 1000]} & \textbf{37 [24, 49]} & \textbf{0.86 \textpm \ 0.06} & \textbf{64 \textpm \ 8.2 \%} \\
\bottomrule
\end{tabular}
\end{table}

On the other hand, the LFM models are, by design, extremely sensitive to the input pose. Indeed, these models achieve state-of-the-art pose selection performance when reranking poses from both docking methods tested. We also see a more dramatic increase in enrichment when using more accurate docked poses (\textsc{DiffDock} versus Uni-Dock), and even state-of-the-art performance when utilizing crystal poses.

We additionally hypothesize that \textsc{DiffDock} gives unrealistic poses for the decoy molecules. This is evidenced in Table~\ref{tab:med_crystal_metrics}, as all models show improved performance discriminating actives with crystal poses from decoys with \textsc{DiffDock} poses than decoys with Uni-Dock poses. Clearly, the models are giving lower scores to the decoys docked by \textsc{DiffDock}. This makes sense when we consider that \textsc{DiffDock} was only trained on binders from the PDB.

With this in mind, we hypothesized that using \textsc{DiffDock} to dock the actives and Uni-Dock to dock the decoys is a better test case for how scoring models perform in the regime where docking models are all around better. Indeed, we see that LFM models are able to perform best in this case. This is consistent with our hypothesis that these models are more pose-sensitive than the others.


Though prospective use of LFM will, of course, use docked rather than crystal poses, we suggest that this pose sensitivity of the LFM models corresponds to the fact that they have learned more generalizable features. Actual binding affinities are highly dependent on the specific intermolecular interactions being made by the protein-ligand complex. Thus, while the LFM models perform as well as the baselines on these relatively ``in-domain'' targets, we hypothesize that LFM will be more reliable when used for more novel targets.

\begin{table}
\centering
\caption{Median metrics for each model when using Uni-Dock or \textsc{DiffDock} for both active and decoy poses.}
\label{tab:med_metrics}
\begin{tabular}{lllllll}
\toprule
Docking method & Model & \EFB{max} & \EFB{1\%} & AUC \\
\midrule

\multirow{3}{*}{Uni-Dock} & Vina & \textbf{9.7 [3.4, 15]} & 1.9 [0.86, 2.7] & 0.54 \textpm \ 0.03 \\
& \textsc{Gnina} & 6.9 [4.2, 14] & \textbf{3.7 [2.1, 4.5]} & \textbf{0.59 \textpm \ 0.02} \\
& LFM & 4.6 [3.3, 14] & 2.0 [1.2, 3.1] & 0.52 \textpm \ 0.02 \\
\midrule

\multirow{3}{*}{\textsc{DiffDock}} & Vina & \textbf{16 [4.7, 25]} & 3.5 [2.0, 4.4] & 0.60 \textpm \ 0.02 \\
& \textsc{Gnina} & 13 [7.7, 27] & 5.8 [3.9, 7.3] & \textbf{0.68 \textpm \ 0.02} \\
& LFM & 14 [7.8, 33] & \textbf{6.2 [3.6, 8.7]} & 0.58 \textpm \ 0.03 \\

\midrule
\multirow{3}{*}{\shortstack[l]{\textsc{DiffDock} (actives) \\ Uni-Dock (decoys)}} & Vina & 1.0 [1.0, 2.6] & 0.35 [0.0, 0.62] & 0.23 \textpm \ 0.02 \\
& \textsc{Gnina} & 7.9 [4.9, 17] & 3.9 [2.3, 5.0] & \textbf{0.65 \textpm \ 0.02} \\
& LFM & \textbf{13 [7.9, 31]} & \textbf{5.8 [2.8, 8.3]} & 0.55 \textpm \ 0.04 \\

\bottomrule
\end{tabular}
\end{table}

LFM inference took an average of 2.5 s on an L40 GPU per molecule, which is fast enough to use in large-scale screening.

Overall, we see that our LFM models demonstrate remarkable early enrichment for three of the six targets. The other three, however, proved more difficult for the method.

\subsection{TBXT virtual screening}

To test how well LFM and the baselines work on novel binding sites in a real-world setting, we ran a virtual screening campaign for the TBXT transcription factor. This is an important and currently undrugged target implicated in chordoma and other cancers \cite{hamilton_development_2017, sharifnia_small-molecule_2019}. We ran a 4 μs MD simulation of TBXT and identified a potential cryptic binding site. Since this is a novel binding site on an already novel target, it represents an incredibly difficult but realistic test case for virtual screening methods. While a couple (weak) binders for this target exist in the PDB, none bind to the site we identified.

We used Uni-Dock to generate binding poses for 1M random compounds from the Enamine REAL set \cite{grygorenko_generating_2020}. We used the baseline models as well as an LFM model trained on 50K datapoints to rescore these compounds and ran ABFE calculations on the top 50 compounds proposed by each method. We used the method proposed by \citet{ding_bayesian_2024} to provide robust uncertainty estimates for each prediction; since these predictions had high variance, we used the upper bound 95\% confidence interval as a conservative estimate of the binding free energy. This can sometimes give unrealistically high ΔG estimates (in general, the true binding ΔG will be less than 0 kcal/mol), but these unrealistic estimates are more often given for poor binders with unstable binding poses anyway (as the current ABFE code assumes a relatively stable binding pose). As a control, we also ran these calculations on 50 random compounds. See Appendix \ref{app:abfe} for more details on the free energy calculations.

While this benchmark is far from perfect, as we are using predicted values rather than experimental binding affinities, we argue that such a benchmark still has value. As mentioned earlier, ABFE calculations are known to rank-order ligands well and are quite effective at discriminating actives from decoys \cite{alibay_evaluating_2022, fu_meta-analysis_2022, aldeghi_accurate_2015}.

We noticed that the top-scoring compounds proposed by the LFM method sometimes had strange poses on the side of the pocket. We reasoned that this could be an artifact of the training procedure, as such poses are unlikely to be representative of ligands in the training set; perhaps a predicted low-energy region existed there by accident. We hypothesized that combining LFM and Vina models would produce better top-scoring compounds by grounding the model and preventing these sort of ``hallucinations''. Thus we also tested the top 50 compounds when we combined the ranks of Vina and LFM scores (and only keeping compounds whose top-scoring Vina and LFM conformer was within 2 \AA). We also tested this same procedure with \textsc{Gnina} and Vina, using the same reasoning. Finally, we tested a combination of all three models.

We also used the recently released Boltz-2 as a baseline \cite{passaro_boltz-2_2025}. Because Boltz-2 takes much longer to run than the other models (around 5 minutes rather than a couple of seconds), we were unable to run it on the entire library of 1M compounds. Instead, we used it to screen 10K compounds. We also ran the Vina and LFM method on these same compounds as a direct comparison. We did not benchmark more models on this 10K library due to the expense of running ABFE calculations; our primary goal was to test if the accuracy of Boltz-2 was enough to compensate for its speed. It's worth noting that we did not benchmark Boltz-2 against the previous well-characterized targets since it was (likely) trained on all of them.

\begin{figure}
\caption{Distribution of predicted binding affinities after running ABFE simulations on the top 50 results from each scoring method.}
\centering
\includegraphics[width=0.8\textwidth]{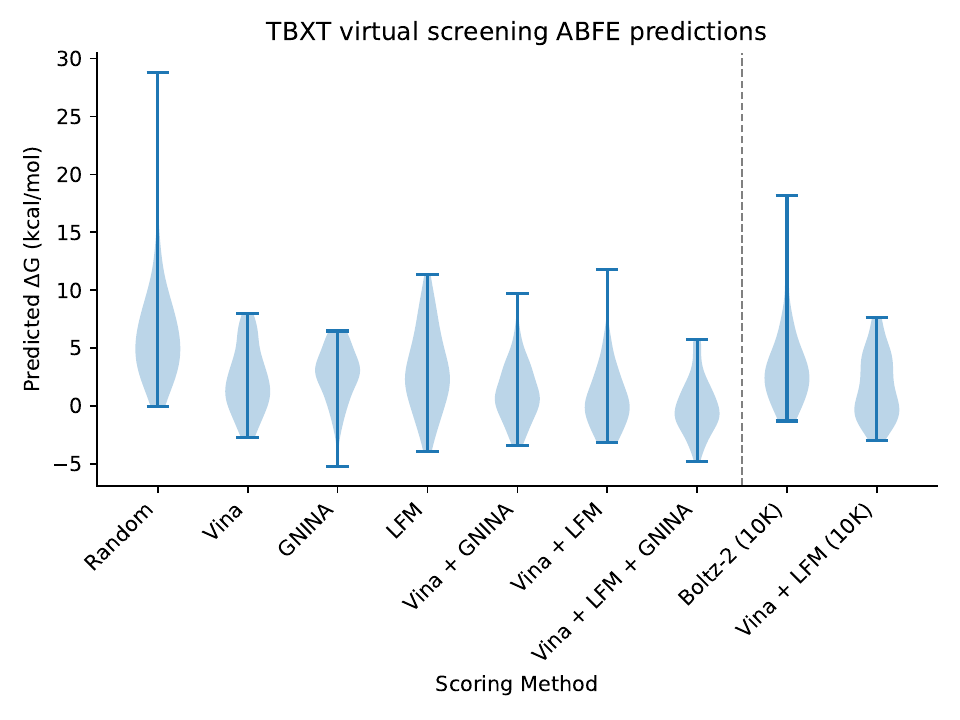}
\label{fig:tbxt_violin}
\end{figure}

The results of this screening benchmark are shown in Figure~\ref{fig:tbxt_violin} and Table \ref{tab:tbxt_metrics}. All models are significantly better than random guessing (as measured by mean predicted ΔG). Curiously, Vina, \textsc{Gnina}, and the LFM model all perform similarly, and ensembling multiple models helps a lot. The combination of all three models performs the best. Though the triple combination is not a statistically significant improvement over the Vina + LFM combination ($p=0.097$), it is over the Vina + \textsc{Gnina} combination ($p=0.008$).

Notably, the Vina + LFM combination is a significant improvement over Boltz-2 when using a 10K library ($p = 0.003$), despite running approximately 100x faster.


\begin{table}
\centering
\caption{TBXT virtual screening results. For each model, we report the mean and median predicted binding affinities (as given by our ABFE calculations). We also return the percentage below 0 kcal/mol.}
\label{tab:tbxt_metrics}

\begin{tabular}{lllll}
\toprule
Library size & Model & Median pred. ΔG $\downarrow$ & Mean pred. ΔG $\downarrow$& $\% < 0$ kcal/mol $\uparrow$ \\
\midrule

\multirow{5}{*}{1M}
&Vina & 1.8 \textpm \ 1.0 & 2.2 \textpm \ 0.77 & 22 \textpm \ 12 \\
&\textsc{Gnina} & 2.8 \textpm \ 0.68 & 2.7 \textpm \ 0.65 & 12 \textpm \ 9.0 \\
&LFM & 2.6 \textpm \ 1.2 & 3.1 \textpm \ 0.99 & 22 \textpm \ 11 \\
&Vina + \textsc{Gnina} & 0.89 \textpm \ 0.70 & 1.2 \textpm \ 0.73 & 34 \textpm \ 13 \\
&Vina + LFM & 0.22 \textpm \ 0.75 & 0.68 \textpm \ 0.81 & 44 \textpm \ 14 \\
&Vina + LFM + \textsc{Gnina} & \textbf{-0.15 \textpm \ 0.65} & \textbf{-0.0027 \textpm \ 0.71} & \textbf{56 \textpm \ 13} \\
\midrule
\multirow{2}{*}{10K}
&Boltz-2 & 2.8 \textpm \ 0.89 & 3.4 \textpm \ 0.98 & 12 \textpm \ 9.0 \\
&Vina + LFM & \textbf{0.79 \textpm \ 1.2} & \textbf{1.6 \textpm \ 0.77} & \textbf{36 \textpm \ 13} \\

\bottomrule
\end{tabular}

\end{table}

\section{Discussion}

We presented ligand force matching, a new approach to protein-ligand scoring that utilizes molecular dynamics simulations to train a per-target neural network. We demonstrated the potential of this method on 6 protein targets, utilizing 100-500 μs of MD simulation data per target. The LFM models achieve a state-of-the-art early enrichment when using the true co-crystal poses for the active compounds, and the performance remains competitive when using docked poses.

Ligand force-matching is a general method for developing virtual screening models for novel targets. The specific workflow used in this paper is just one possible variant of the method, and there are many possible improvements that could further boost model performance. For instance, there are many possible neural network architectures to try beyond the equivariant transformer used in this work. Of particular interest is the Gaussian process approach used by \citet{vandermause_--fly_2020}, which could enable robust uncertainty estimation for our predictions.

\iflong
On the data generation side, there are many possible things to try. For instance, the current workflow presented only uses a single protonation state for the protein when generating data, but, in reality, histidines in the binding site often change protonation state based on the bound ligand A more careful approach would be to assign protonation state separately for each pose in the training set before MD simulation, or to do more advanced MD simulations at constant pH \cite{donnini_constant_2011, chen_constant-ph_2015}. Additionally, there could be slow timescale movements that the 4 ns simulations could not capture. Longer simulations during data generation (or enhanced sampling methods) could alleviate this. Finally, there are inherent inaccuracies in the molecular mechanics (MM) force field. New neural network potentials (NNPs) have recently shown great success at quickly approximating quantum mechanical energies, and even in improving protein-ligand binding affinity predictions \cite{smith_ani-1_2017, duignan_potential_2024, sabanes_zariquiey_enhancing_2024}. Current NNPs are still much slower than MM force fields, but we could use an NNP to reweigh the forces from an MM simulation.
\fi

The most important limitation of this method is the requirement for large-scale molecular dynamics simulations for each target studied. While this is generally worth the cost (high-throughput screening still costs orders of magnitude more money), future work should be done in creating MD datasets across the proteome to train a general-purpose network.

While we focused on the application of this method to scoring, a similar method could also be used for docking, as a ligand's pose in the binding site should be the minimum of the free energy surface. The LFM models presented already outperform the baselines in terms of reranking poses from Uni-Dock and \textsc{DiffDock}. Going forward, we intend to train smaller versions of the network (similar to the simple energy functions used by Vina or Glide) that would allow for quickly minimizing the predicted free energy. This would allow for predicting poses that require greater receptor rearrangement, a task at which current docking methods struggle.

We intend to use this method for prospective hit discovery for novel targets, and we hope that other researchers do the same. There is also much work to be done in testing the limits of this method. How much binding site flexibility can it handle? How well does it work for more difficult target classes, such as intrinsically disordered proteins, protein-protein interactions, or even RNA targets?

In conclusion, we believe that ligand force matching has the potential to transform early-stage drug discovery for novel targets. We hope that future researchers continue to test and refine this method so it can achieve its potential.

We release the code for the dataset generation, model training, and model benchmarking at \url{https://github.com/molecularmodelinglab/lfm} under the MIT license.

\iflong

\begin{ack}
The authors thank Dr. David Koes, Dr. Yinglong Miao, Enes Kelestemur, Rishabh Dey, Kushal Koirala, and Nyssa Tucker for insightful comments and discussion.

\end{ack}
\fi

\bibliography{references}

\appendix

\section{Technical Appendices and Supplementary Material}

\subsection{\label{app:datagen}Dataset generation}

For BRD4, the first target that the LFM workflow was tested on, we generated the ligand poses for the dataset slightly differently. Instead of randomly initializing the ligand in the binding site, we used QuickVina \cite{alhossary_fast_2015} to initialize the pose in a random protein conformation from the apo trajectory. We then generated a random translation that took the ligand outside the binding site, and a random rotation. We then produced the final translation as a uniformly random interpolation between the docked and random poses. Finally, we froze the ligand in this pose and alchemically added it to the random apo protein frame. The intuition was to sample the full space of ligand poses while sampling more around ``good'' poses given by QuickVina, similar to how datapoints are generated in flow-matching models. While this strategy worked well for BRD4, it quickly became apparent that it would not work well in general. The binding sites for the other proteins in the benchmarking set tended to close during the apo trajectory, meaning that QuickVina was rarely able to find ``reasonable'' ligand poses. Hence we elected to go for the less biased data generation approach presented above.

For CDK2 and HSP90, we used a slightly different ligand restraint algorithm when generating the data. Specifically, we restrained the distance between the ligand center of mass and the protein center of mass to be the same as the one when the ligand was randomly positioned. We changed it to a full restraint on the ligand center of mass to the original ligand center of mass to keep the ligand closer to the ``exit path,'' thus (hopefully) being more data efficient.

\subsection{\label{app:model}Model architecture}

Each model was a 3-layer Equivariant Transformer architecture as implemented in TorchMDNet \cite{tholke_torchmd-net_2022} with a hidden dimension of 128. To produce input atom embeddings for the model, we first generated a ``receptor embedding'' from 3 interpolated cubice spline grids with hidden dimension [64, 256, 512] and spacing [2 \AA\, 4 \AA\, 8 \AA]. The grid was sized so that it included all the residues in the protein binding site and the ``exit point'' for the binding site. We concatenated these interpolated embedding and applied a Dropout layer \cite{srivastava_dropout_2014} before producing a 128-dimensional hidden vector. We additionally generated a ``ligand embedding'' as a linear layer mapping the atom nonbonded parameters (partial charge $q$ and Lennard-Jones $\epsilon$ and $\sigma$) to a 128-dimensional hidden vector before applying Dropout. Finally, we produced a ``both embedding'' by multiplying both the embeddings. The final atom embedding was a linear layer and Dropout over the concatenated ligand, receptor, and both embeddings to produce a 128-dimensional hidden vector. We used a SiLU activation function for all the layers described \cite{hendrycks_gaussian_2023}, and a dropout rate of 0.2 for all the Dropout layers.

\subsection{\label{app:vs}BRD4 virtual screen}

To run the BRD4 virtual screen, we used Uni-Dock to dock the entire in-house library before rescoring the ligands with the LFM model (unlike the results presented in this paper, we minimized the docked poses with the LFM model before rescoring). We nominated the top ten virtual hits with less than 6 rotatable bonds, ensuring that no hit with within 0.3 Tanimoto similarity to any other hit.

\subsection{\label{app:rigid} Rigid-body restraining force}

When generating the dataset to train the LFM models, we ensure that the receptor remains in the reference frame defined by the pocket $\alpha$ carbons by the following force:

\[ F_i = \overline{F} + \frac{\overline{T}_i \times x_i}{|x|^2}\]

Here $\overline{F}$ and $\overline{T}$ are the mean force and torque over the pocket $\alpha$ carbons for the unprojected harmonic force $\hat{F_i} = -2k(x_i-\hat{x}_i)$, where $\hat{x}$ are the original pocket $\alpha$ carbon positions. In this work, we used $k$ = \SI{e4}{\kilo\joule\per\mol\per\nano\metre\squared}. 

\subsection{Model scoring}
Before using the LFM models to score docked poses, we minimize the intramolecular energy of the ligand (with harmonic restraints to the original docked coordinates with k=\SI{5}{\kilo\joule\per\mol\per\nano\metre\squared}). We do this because the docked poses may have bond lengths and angles that are out-of-domain relative to the poses that the model has been trained on.

\subsection{\label{app:abfe} Free energy calculations}
We used the double-decoupling ABFE method \cite{gilson_statistical-thermodynamic_1997} with nonequilibrium switching to estimate absolute binding free energies. We used a vacuum intermediate ligand state that was harmonically restrained to its binding pose (taken after a 5 ns simulation) with k=\SI{10}{\kilo\joule\per\mol\per\nano\metre\squared}). We equilibrated the ligand-only, apo, and holo simulations for 10 ns before sampling frames every 500 ps for the switching. We used 30 2 ns forward and backward switches when decoupling the ligand from the receptor, and 60 2 ns forward and backward switches when decoupling the ligand from solvent.

\end{document}